\documentstyle[pre,aps,epsf]{revtex}

\begin{document}

\title{FT-194-1980, August}

\vspace{1.0cm}
\author{\bf \LARGE Hartree--Fock--Bogoliubov Approximation for Finite Systems }

\vspace{0.75cm}

\address{{\large{\bf {\rm Aurel Bulgac}}}
\footnote{Present address: Department of Physics, University of
Washington, Seattle, WA 98195--1560, USA}\\ 
Central Institute of Physics \\
Institute for Physics and Nuclear Engineering \\
Bucharest, POB MG--6, ROMANIA }

\maketitle

\begin{abstract}

{\it
The original  preprint was never published, due to various circumstances
beyond the author's control. The initial preprint
has been known to a number of people, the results have been used
in published literature and the preprint is still been referred to.
The present text has been edited somewhat and a number of typos have been
corrected; however, no significant changes have been incorporated into
the text. No formulas have been added or deleted and the figures have
been redrawn. A scanned copy of the original preprint in jpeg format 
can be found at \\
$ http://www.phys.washington.edu/~bulgac/publications.html $ 

The author thanks F.M. Edwards for helping to edit the text and Yongle
Yu for bringing to his attention a number of typos.}

\vspace{0.75cm}

Some general features of the spectrum of the Hartree--Fock--Bogoliubov
equations are examined. Special attention is paid to the asymptotic
behavior of the single quasiparticle wave functions (s.qp.w.fs.),
matter density distribution and density of the pair condensate. It is
shown that due to the coupling between hole and particle states,the
deeply bound hole states acquire a width and have to be
treated as continuum states. The proper normalization of the
s.qp.w.fs. is discussed.

\end{abstract}
\newpage

\section{Introduction}

The Hartree--Fock--Bogoliubov (HFB) approximation was outlined more
than twenty years ago \cite{NNB} for infinite systems and almost
immediately was introduced in nuclear physics \cite{STB}. In the
case of infinite systems the HFB procedure is well studied and the
character of the wave--functions (w.f.)  is well understood. However,
in the case of finite systems (nuclei), things are not so
clear. The HFB equations in the case of nuclei are meaningful provided
the boundary conditions for the single quasiparticle (s.qp.) w.fs.
are correctly formulated in order to describe a genuine finite
state. The goal of the present paper is to provide the correct formulation
of the HFB approximation in the case of finite systems.

It is well known that pairing correlations always appear whenever a pole
(i.e. a bound state) is present in the two-body Green function of the
many-body system \cite{ABM}. In this case, corrections to the
single particle (s.p.)  Green function of the type
\begin{figure}[h]
\begin{center}
\epsfxsize=3.5cm
\centerline{\epsffile{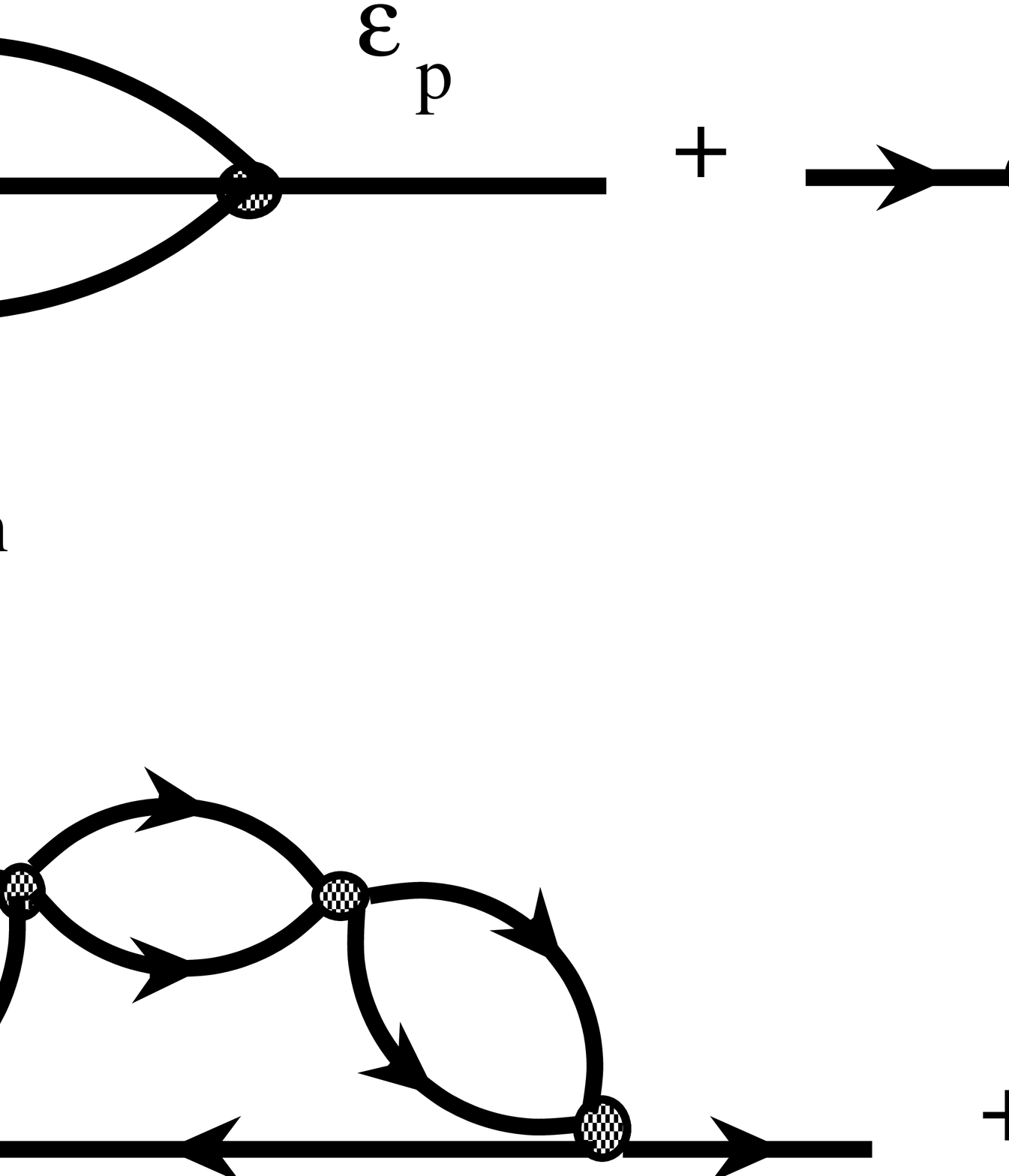}} 
\end{center}
\end{figure}
\noindent give rise to diagrams of the type
\begin{figure}[h]
\begin{center}
\epsfxsize=3.5cm
\centerline{\epsffile{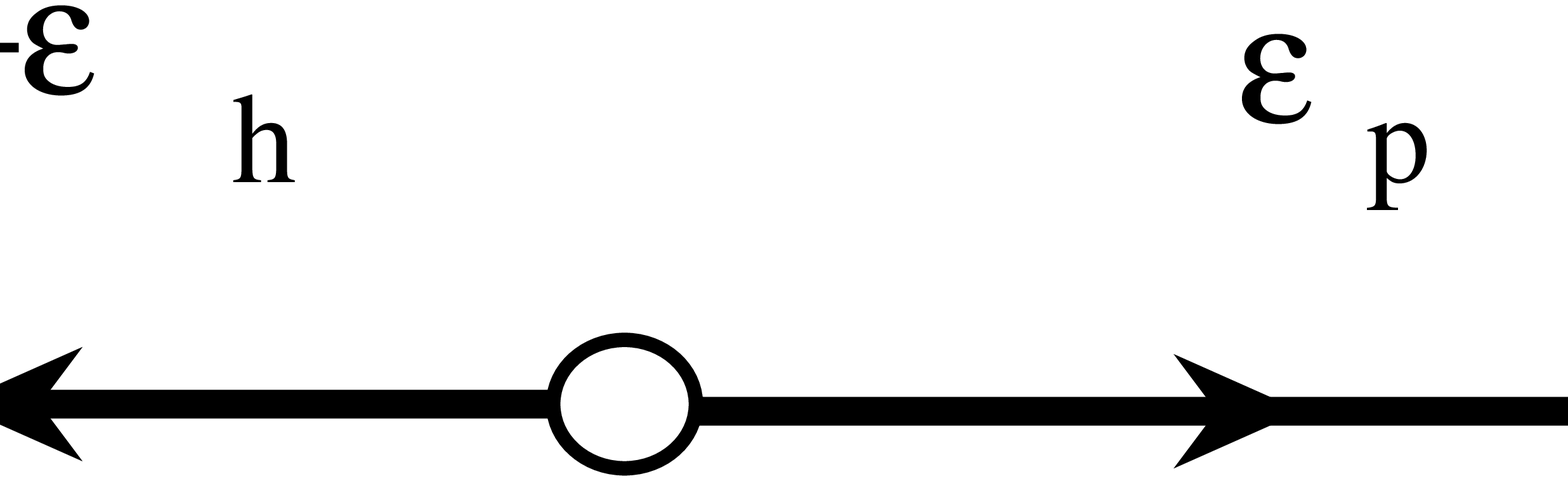}}
\end{center}
\end{figure}
\noindent when only the contribution of the pole is taken into account
\cite{ABM}. Here $2 \lambda$ stands for the energy of the two--particle
bound state and $\varepsilon_p$ and $\varepsilon_h$ are the particle
and hole energies respectively. The process represented by diagram (1)
leads to a mixing between particle and hole states and as a result to a
smearing of the Fermi surface \cite{ABM,JGV,MB}. This mixing has the
special feature that a hole (particle) can transform into a particle
(hole), due to the presence of the pair condensate, provided their
energies are related by the energy conservation law
\[ \varepsilon_p + \varepsilon_h = 2 \lambda . \] 
Whenever the energy of the hole state is less than $2 \lambda$ the
corresponding particle state to which the hole state is coupled lies
in continuum. This situation formally resembles the case of an
electron in a very strong field \cite{ABM2} or the case of a bound
state embedded in continuum \cite{CM}. Consequently, a sufficiently
deep hole state becomes unstable with respect to decay into a
particle state by an interaction with the pair condensate. The same
thing happens in the case of a deep hole decaying into a
less deeply bound hole, with the excitation of a phonon, which can
further decay by particle emission. This process is described by the diagram
\begin{figure}
\begin{center}
\epsfxsize=3.5cm
\centerline{\epsffile{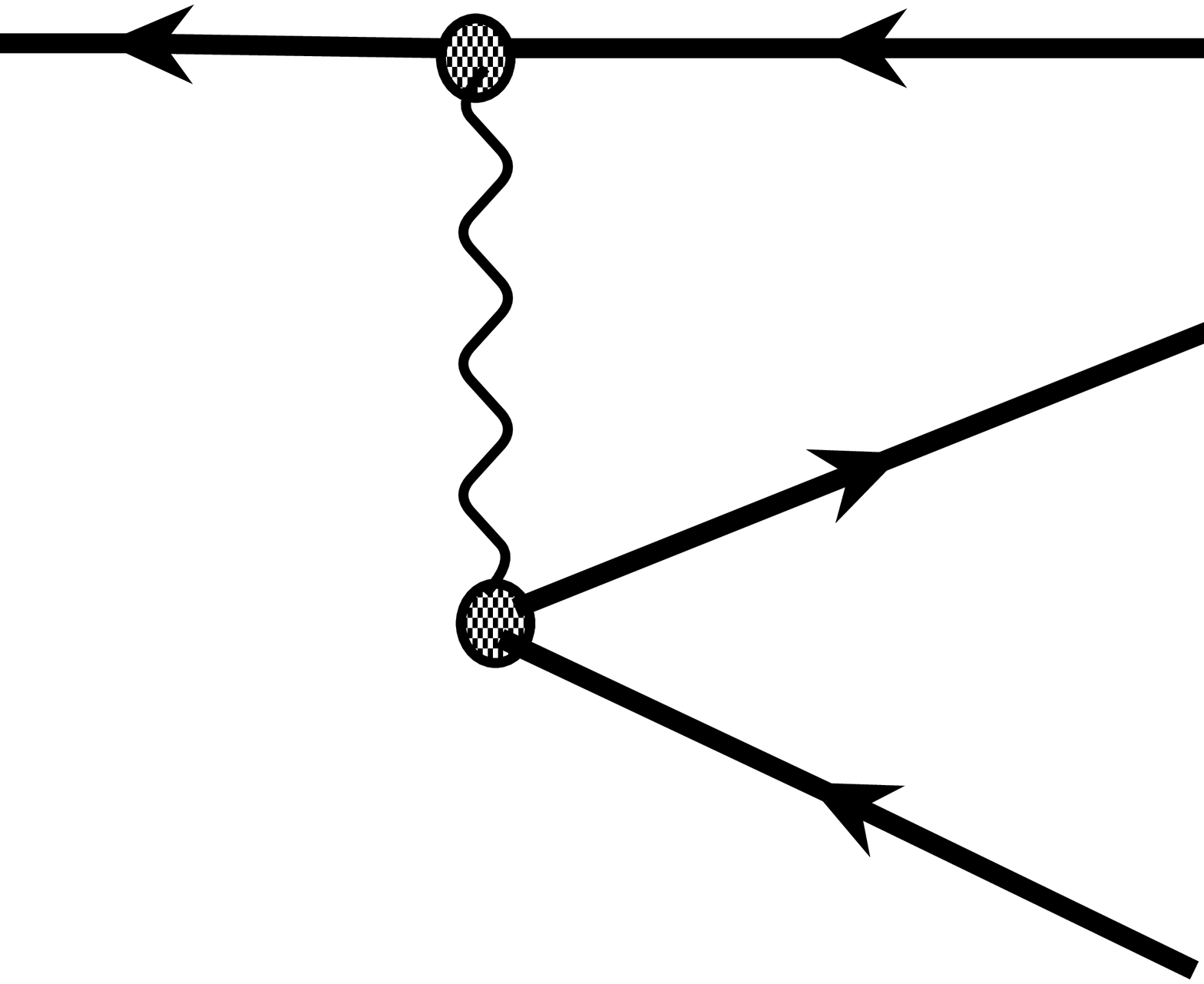}}
\end{center}
\end{figure}
\noindent{where the wavy line represents a phonon.}

The physical situation is thus not new. New is the fact that due
to the presence of the pair condensate, some hole states acquire a
width and therefore the s.qp.w.fs. can no longer be treated as
corresponding to bound states, as has hitherto been done, but as
continuum states. How to introduce this property of
the s.qp.w.fs. into the HFB approximation for finite systems and
therefore how to define the boundary conditions correctly is the main
goal of the present discussion.

\section{HFB equations for finite systems}

In this section the well--known HFB equations will be derived for the sake
of completeness. The emphasis will be on the correct definition
of s.qp.w.fs. so as to describe genuine finite systems, i.e. systems
with finite matter distribution. The forces between particles will not
be specified, except for some  general properties, like the
finite range.

By analogy with the usual HF approximation, the HFB ground state w.f.
$\mid 0\rangle $ is defined as the vacuum for the fermi quasiparticles
\cite{NNB,STB,JGV,MB}
\begin{equation}
\setcounter{equation}{2}
\alpha_i \mid 0 \rangle =0 \;\; , \quad \langle 0\mid 0\rangle=1 ,
\end{equation}
where
\begin{eqnarray}
\alpha_i &=&\int dx [u_i^* (x)_i \psi (x)+ v_i (x)\psi^\dagger  (x)] , 
\\
\alpha_i^\dagger  &=& \int dx [ v_i^* (x) \psi (x)+ u_i (x)\psi^\dagger (x) ]
,\nonumber 
\end{eqnarray}
and $\psi(x)$ and $\psi^\dagger (x)$ stand for field operators for 
annihilation and creation of a particle with space--spin coordinates 
$x=(\vec{r},\sigma)$, which satisfy the usual anticommutation relations
\begin{eqnarray} 
& &\{ \psi (x) , \psi^\dagger  (y) \} = \delta (x-y) ,\\
& &\{ \psi (x) , \psi (y) \} = ( \psi^\dagger (x) , \psi^\dagger (y) ) 
=0. \nonumber 
\end{eqnarray}
By requiring that $\alpha_i$ and $\alpha^\dagger_i$ represent fermion
operators, i.e.
\begin{eqnarray}
& &\{ \alpha_i , \alpha^\dagger_j \} = \delta_{ij},  \nonumber \\
& &\{ \alpha_i , \alpha_j \} = \{ \alpha^\dagger_i , \alpha^\dagger_j \}
=0 ,\nonumber 
\end{eqnarray}
one easily obtains the relations
\begin{eqnarray}
& &\int [ u_i^* (x) u_j (x) + v_i (x) v_j^* (x) ] dx = \delta_{ij}  ,\\
& &\int [ u_i^* (x) v_j (x) + v_i (x) u_j^* (x) ] dx = 0 ,\nonumber 
\end{eqnarray}
\begin{eqnarray}
& &\sum_i u_i (x) u_i^* (y) + v_i (x) v_i^* (y) = \delta (x-y)  ,\\
& &\sum_i u_i (x) v_i (y) + v_i (x) u_i (y) = 0. \nonumber
\end{eqnarray}
The constraints (5--6) ensure the unitary character of the
transformations (3).

The total energy of the many--body system and the mean number of
particles are
\[ {\cal E} = \langle 0 \mid \hat{H} \mid 0 \rangle \]
and
\[ N = \langle 0 \mid \hat{N} \mid 0 \rangle , \]
where $\hat{H}$ and $\hat{N}= \int dx \psi^\dagger (x) \psi (x)$ 
stand for the hamiltonian and the number operator in the second 
quantization representation.

The mean values for the energy and for the particle number can be expressed
through the densities
\begin{equation}
\rho (x,y) = \langle 0 \mid \psi^+ (y) \psi (x) \mid 0 \rangle = 
\rho ^* (y,x) = \sum_i v_i (x) v_i^* (y),
\end{equation}
\begin{equation}
\Phi (x,y) = \langle 0 \mid \psi (y) \psi (x) \mid 0 \rangle = - \Phi (y,x) 
= \sum_i v_i (x) u_i (y),
\end{equation}
in the following way
\begin{equation}
{\cal E} = {\rm Tr} (  T\rho ) + \frac{1}{2} {\rm Tr} 
( V\rho\rho)_a + \frac{1}{2} {\rm Tr} (V\Phi\Phi ^*),
\end{equation}
\begin{equation}
{\rm N} = {\rm Tr} (\rho),
\end{equation}
if only two--body interactions are present. $T$ and $V$ stand for the
kinetic energy and the two--body interaction respectively. A shorthand 
notation was used for the traces in Rels. (9) and (10).

The HFB equations for the two component s.qp.w.fs.  $\{ v_i (x),u_i^* (x)
\}$ are derived from the stationarity condition of the total energy (9)
under the subsidiary conditions (10) and (5). These equations are
\begin{equation}
\int h (x,y) v_i (y) dy - \lambda v_i (x) + \int \Delta (x,y) 
u_i^* (y) dy = E_i v_i (x)
\end{equation}
\begin{equation}
\int \Delta^\dagger (x,y) v_i (y) dy - \int h^* (x,y) u_i^* (y) dy 
+ \lambda u_i^* (x) = E_i u_i (x)
\end{equation}
where
\[ \lambda = \frac{\delta {\cal E}}{\delta N} < 0 \]
is the chemical potential, and
\[ h(x,y) = \frac{\delta  {\cal E}}{\delta \rho (x,y)} = h^* (y,x) \]
and
\[ \Delta (x,y) = \frac{\delta {\cal E}}{\delta \Phi^* (x,y)} 
= - \Delta (y,x) \]
are the s.p. hamiltonian and the pairing field respectively and $E_i$
stands for the s.qp. energies.

When performing the summations in Rels. (7) and (8) one must include
only those solutions of the nonlinear system (11--12) with $E_i<0$,
which define the operators $\alpha_i^\dagger$. The solutions with
$E_i>0$ correspond to operators $\alpha ^\dagger _i$.

Let us analyse in more detail these equations in the case of a
finite system.  The problem which arises is the meaning of the
normalization condition (5); namely, if the right hand side of
Rel. (5) should be a $\delta$--function or a Kronecker symbol, as is
usually the case \cite{STB,MB,HJM,DG}, in complete analogy with the HF
approximation.

A finite system is characterized by a finite matter distribution and
therefore by a finite range of the s.p. field (except in the case of
the Coulomb interaction). Naturally, being determined by the matter
distribution, the pairing field must also have a finite range (an
infinite range of the pairing field can only occur if the system under
consideration is unstable with respect to two particle decay.) The
problem is determining the mechanism which leads to finite matter
distribution when s.qp. states no longer have the character of
discrete states, as was discussed in the Introduction.

When pairing is turned off (i.e. $\Delta \equiv 0$) the spectrum of
Eqs. (11--12) looks like the one show in Fig. 1. The left hand
side corresponds to the spectrum of Eq. (11), while the right hand
side corresponds to Eq. (12).

\begin{figure}[h,t,b]
\begin{center}
\epsfxsize=7.5cm
\centerline{\epsffile{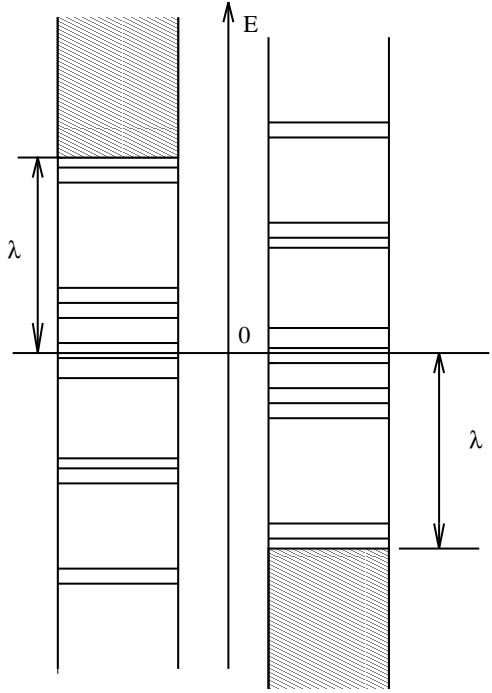}}
\end{center}
\caption{The spectra of Eqs. (11--12) when the pairing field is turned off
($\Delta =\equiv 0$). The left hand side corresponds to the
spectrum of Eq. (11), while the right hand side corresponds to Eq. (12)
. The hatched regions represent the continuum and the
lines correspond to discrete states.}
\label{fig1}
\end{figure}

Upon turning the pairing on, the two spectra mix and the discrete
states with energies outside the interval
\begin{equation}
\lambda < E < - \lambda
\end{equation}
will lie in the continuum. Only the states with energy within the interval
(13) will preserve the bound state character.

The continuum part of the spectrum for $E<\lambda$ and $E>-\lambda$ can
disappear only if the s.p. potential is finite and positive at
infinity as in the case of a harmonic oscillator.\footnote{The
discrete character of the spectrum for a limited number of states with
$E<\lambda (E>-\lambda)$ can also be preserved if $[h,\Delta]=0$, a
condition which is not fulfilled in fact. This condition is
satisfied in the case of constant pairing approximation $( \Delta
\equiv {\rm const.}  )$, an approximation which leads to an unphysical
density distribution.} There is no physical reason for this to be true
in the case of finite systems.

A glance at Fig.1. reveals why
pairing correlations lead to a significant increase of the level
density in the vicinity of the Fermi surface. Pairing 
appears when the chemical potential has a value within a
shell, as illustrated in Fig. 1.  In such a case, the total number of
hole (the left hand side of Fig. 1) and particle (the right hand
side of Fig. 1) states, which is practically equal to the number
of s.qp. states with pairing included, is almost double the number of
s.p. states in HF approximation.

In order to determine 
what happens in the case of a bound state with an energy $E$, when pairing
is turned on, we shall use perturbation
theory. For the sake of simplicity, the nucleus will be assumed to be
spherical and the pairing field real $\Delta \equiv \Delta^*$.  Also,
the spin and angular variables are assumed to be already separated
from Eqs. (11) and (12) and the corresponding geometrical factors included
in the definition of the single particle fields. 
From Eq. (11) one easily obtains in the
vicinity of a bound state the relation
\begin{equation}
v_E(r)= \left ( \frac{1}{E+\lambda-h} \Delta u_E\right ) (r) 
\cong \phi_0 (r) 
\frac{\langle \phi_0 \mid \Delta \mid u_E \rangle}{E-E_0} 
= n^{1/2} (E) \phi_0 (r) , 
\end{equation}
where
\[ (h - \lambda - E_0) \phi_0 = 0 \;\; , \;\; \langle \phi_0 \mid 
\phi_0 \rangle = 1 . \]
The other component of the single quasi--particle wave function becomes then
\begin{equation}
u_E (r) = C (E) u_{0E} (r) - n^{1/2} (E) 
\left ( \frac{1}{-E+ \lambda -h} 
\Delta \phi_0\right ) (r) ,
\end{equation}
where
\begin{equation}
(h- \lambda +E) u_{0E} =0 , \quad \langle  u_{0E} \mid u_{0E'} \rangle = 
\delta (E-E^\prime)
\end{equation}
and $C(E)$ is a normalization constant. From Rels. (14) and (15) 
one obtains
\[ n^{1/2} (E) = \frac{C(E) \langle \phi_0 \mid \Delta \mid
u_{0E}\rangle } 
{E-E_0 + \langle \phi_0 \mid \Delta \frac{1}{-E+\lambda -h}
\Delta \mid \phi_0 \rangle} . \] 

In order to determine the normalization constant $C(E)$, we shall use
the representation of the Green function through regular and irregular
solutions of Eq. (16)
\[ G(r,r^\prime ,-E+\lambda) = 
\left ( \frac{1}{-E+\lambda -h} \right ) (r,r^\prime)
= \frac{2m}{\hbar^2} \frac{u_{0E} (r_<) \chi _E (r_>)}{W
(u_{0E}, \chi _E)} \]
where $W(u_{0E},\chi _E)$ stands for the Wronskian
\[ W ( u_{0E}, \chi_E ) = -\frac{2m}{\pi \hbar^2} .\]
The asymptotic behavior for the regular $u_{0E}$ and irregular $\chi _E$
solutions is given by
\begin{eqnarray}
& &u_{0E} (r) \cong \sqrt{\frac{2m}{\hbar^2 \pi k}} 
\sin (kr +\delta_{0E})  , \nonumber\\
& &\chi _E (r) \cong \sqrt{\frac{2m}{\hbar^2 \pi k}} 
\cos (kr +\delta_{0E}) , \nonumber
\end{eqnarray}
where $k^2 = 2m (-E + \lambda ) /\hbar^2$.  The asymptotic behavior of
the $u$--component is
\[ u_E (r) \cong \sqrt{\frac{2m}{\hbar^2 \pi k}} 
[ C(E) \sin (kr+ \delta_{0E})+ \pi n^{1/2} (E) \langle \phi_0 \mid \Delta 
\mid u_{0E} \rangle \cos (kr+ \delta_{0E}) ],  \]
and therefore
\begin{equation}
C^2 (E) = \frac{[ E-E_0 + 
\langle \phi_0 \mid \Delta \frac{1}{-E+\lambda -h}
\Delta \mid \phi_0 \rangle  ]^2}{[ E-E_0 + < \phi_0 \mid \Delta 
\frac{1}{-E + \lambda -h} \Delta \mid \phi_0 > ]^2 + \pi^2 \mid 
< \phi_0 \mid \Delta \mid u_{0E} > \mid^4}.
\end{equation}
Consequently, the normalized solutions are
\begin{equation}
v_E (r) = n^{1/2} (E) \phi_0 (r)
\end{equation}
\begin{equation}
u_E (r) = C(E) u_{0E} (r) - n^{1/2} (E) 
\left (  \frac{1}{-E + \lambda -h} \Delta \phi_0\right )  (r) \; ,
\end{equation}
where
\begin{eqnarray}
n(E) &=& \frac{1}{\pi} \quad \frac{\frac{1}{2} \Gamma (E)}{(E- E_0
- \delta E)^2 + \frac{1}{4} \Gamma^2 (E)} \; , \\
\Gamma (E) &=& 2\pi | \langle  \phi_0 \mid \Delta \mid u_{0E} \rangle
|^2 = 2\pi \frac{\mid \langle  \phi_0 \mid [ h,\Delta ]
\mid u_{0E} \rangle  \mid^2}{(E+E_0)^2}  ,\nonumber \\
\delta E & =& - \langle \phi_0 \mid \Delta \frac{1}{-E+\lambda -h}
\Delta \mid \phi_0 \rangle \nonumber
\end{eqnarray}
and
\begin{equation}
\int_\infty^\infty n(E) dE =1.
\end{equation}
The matrix elements in the above relations can be calculated without
any significant loss of accuracy for $E=E_0$.

As one can observe, due to the coupling with the continuum, the bound
state spreads over the entire spectrum (see Rel. (21)). Now the quantity
$n(E)$ has thus to be interpreted as the occupation number probability
density over a unit energy interval.

The solution is formally equivalent to the solution of a coupled
channel problem with a bound state embedded in the continuum\cite{CM}.
It displays a well--defined resonant character with a width $\Gamma
(E_0)$ and a shift $\delta E$.  The case of a resonant state can be
treated in a similar way.

Far from the resonance energy $E_0 $, the amplitude of the
$u$--component is very small, while the $u$--component is practically
equal to the unperturbed solution $u_{0E}$. In the vicinity of the
resonance  $E_0$, the amplitude of the $u$--component increases significantly 
inside the potential well (see
Rels. (18), (20)), while the phase of the
$u$--component changes by $\pi$. The phase of the $u$--component is
\[ \delta_E =\delta_{0E} -\arctan \frac{\Gamma (E)}{2(E-E_0 - \delta E)} \]
The solution just described is characterized by the fact that the
$v$--component is square integrable, even though the single
quasi--particle wave function solution $( v_E (x), u^*_E (x) ) $
represents a continuum state. The question is: does this feature
hold true for the selfconsistent solution of the HFB equations?

The density distribution $\rho$ satisfies Rel. (10) (i.e. the diagonal
part of $\rho$ is integrable), while the density of the pair condensate
satisfies the condition
\begin{equation}
\int \mid \Phi (x,y) \mid^2 = Tr [\rho (1-\rho )] < N,
\end{equation}
which can be easily derived by means of Rels. (5--6)
\cite{NNB,JGV,MB,HJM}. Therefore, one can expect that $\rho$ and
$\Phi$ both fall down quickly enough outside the system. The
density distribution $\rho$ determines the asymptotic behavior of the
single particle selfconsistent potential $U$, while the density $\Phi$
defines the pairing potential $\Delta = V\Phi/2$.

The nonlocality of these potentials is governed by the range of the
two--body interaction $V$, assumed to be finite.

We shall now show that the following  asymptotic behaviors take place
\begin{equation}
U \sim \rho \sim {\cal{O}}\left ( \exp 
\left [ 2\sqrt{\frac{2m \mid\lambda\mid}{\hbar^2} } r\right ]\right ) 
\end{equation}
and
\begin{equation}
\Delta\sim\Phi\sim {\cal{O}} 
\left ( \exp \left [ \sqrt{2 \frac{2m
\mid\lambda\mid}{\hbar^2}}r \right  ] \right ),
\end{equation}
when $\bbox{R} = (\bbox{r} + \bbox{r}^\prime )/2$ tends to infinity
and $\bbox{s} = \bbox{r}-\bbox{r}^\prime$ remains finite (practically
of the order of the range of $V$).  (In the above Rels. (23--24) the
symbol ${\cal{O}}$ means that the quantities on the left hand side
behave like the corresponding arguments of the ${\cal{O}}$--function.)
As one can observe, the pairing field $\Delta$ has a longer tail than
the s.p. selfconsistent field $U$.

Using Rels. (23) and (24) one can show, using HFB Eqs. (11--12), that the
$v$-- and $u$--components behave asymptotically as
\begin{equation}
v(r) \sim \exp \left (
- \sqrt{\frac{2m}{\hbar^2} \mid E + \lambda \mid} r\right )   ,
\end{equation}
\begin{equation}
u(r) \sim \exp \left ( - \sqrt{\frac{2m}{\hbar^2} \mid E- \lambda
\mid} r\right ) ,
\end{equation}
if
\begin{equation}
\lambda < E < 0 ,
\end{equation}
and
\begin{equation}
v(r) \sim \exp\left (- \sqrt{\frac{2m}{\hbar^2} \mid 2\lambda\mid}
r\right ),
\end{equation}
\begin{equation}
u(r) \sim {\cal{O}}(1),
\end{equation}
if
\begin{equation}
E < \lambda .
\end{equation}

Outside the potential well, for energies in the interval (27), the two
Eqs. (11--12) decouple and the corresponding asymptotic behavior of the
$v$-- and $u$--components of the single quasi--particle wave function
is determined by the ``energies'' $\lambda +E$ and $\lambda -E$,
respectively.  For energies in the interval (30,) the asymptotic
behavior of the $v$--component is governed by the inhomogeneous part
of the Eq. (11) (i.e. by the term $\Delta u^*$), which cannot be
neglected in this case, as was possible for energies in the interval (27). On
the other hand, the term $\Delta^\dagger v$ falls down exponentially
and does not influence the asymptotic behavior of the $u$--component.

The asymptotic behavior of the $u$--component is fully determined by
the ``energy'' $- E + \lambda$, which is positive in the interval (30).

Now, if one takes into account the definitions of densities $\rho$ and
$\Phi$ (Rels. (7) and (8) respectively), one can easily notice that the
asymptotic behavior of the $v$-- and $u$--components (see Rels. (25--30))
completely agrees with the asymptotic behaviors (23) and (24). This
means that the corresponding asymptotic behaviors are selfconsistent.
It is physically natural to expect that the asymptotic behavior of the density
$\rho$ is controlled by the chemical potential $\lambda$, i.e. by the energy of the
least bound particle. The density of the pair condensate $\Phi$ can be
interpreted as the wave function of a bound state of two interacting
particles in an external field with energy $2\lambda$. Using HFB
Eqs. (11) and (12) and the definition (8), one can show that the
density $\Phi$ satisfies the equation \cite{NNB}
\begin{equation}
(h(1)+h(2)+v(1,2) - 2\lambda) \Phi (1,2) =
\end{equation}
\begin{center}{\it other terms negligible outside the system
}.\end{center}

\noindent{From this equation it follows that at large distances $\Phi$ behaves as}
\[ \exp \left (-\alpha r_1 - \beta r_2\right ), \]
where
\[ -( \alpha^2 + \beta^2 ) = \frac{2m}{\hbar^2} 2 \lambda ,  \]
which also agrees with Rel. (24).

Strictly speaking, this asymptote is correct only outside the range of
$V$. If,however, one takes into account the fact that a
system of two identical nucleons does not have bound states, the
asymptote is valid everywhere outside the s.p. potential well.

Summing up, in the normalization conditions (5) for the single
quasi-particle wave functions, the right hand side has to be
interpreted as a Kronecker symbol if $\lambda < E < -\lambda$ and as a
Dirac $\delta$--function if $\mid E \mid > \mid \lambda \mid$ (as it is
well known \cite{NNB,STB,ABM,JGV,MB,HJM} the system (11--12) has the
property that if $\{ v_i, u^*_i, E_i \}$ is a solution, then $\{ u_i,
v^*_i , -E_i \}$ is a solution as well). Furthermore, for $E < 0$ the
$v$--component of the single quasi--particle wave function is always
square integrable and its norm has to be interpreted as the occupation
number probability.  On the other hand, the relation
\[ 1 - n_i = 1 - \int \mid v_i (x) \mid^2 dx = \int \mid u_i (x) \mid^2 dx \]
is valid only for $\lambda < E < - \lambda$. 
If the energy $E$ is outside this interval, the integral $\int |v_i(x)|^2dx$ 
should be interpreted as an occupation number probability density per unit 
energy interval.

\section{Some simple examples}

This section is devoted to some simple examples of HFB equations, which
although somewhat unrealistic, lead to a better
understanding of various issues arising while solving
Eqs. (11--12).

\subsection{Constant pairing approximation $\Delta \equiv$ constant}

This approximation is also known as the BCS approximation \cite{JB}.  The
s.qp.w.f. in this case is
\begin{eqnarray}
v (r) &=& n^{1/2}_{BCS} \phi (r) , \nonumber \\
u (r) &=& \sqrt{1 - n_{BCS}} \phi (r), \nonumber
\end{eqnarray}
where
\begin{equation}
(h- \epsilon ) \phi = 0  , \quad 
n_{BCS} = \frac{1}{2} \left ( 1 - \frac{\epsilon - 
\lambda}{\sqrt{\epsilon - \lambda)^2 + \Delta^2}} \right ) \nonumber
\end{equation}
and
\[ E = - \sqrt{(\epsilon -\lambda)^2 + \Delta^2}. \]

Usually, the pairing field $\Delta \equiv$ const is taken to be different
from zero in a limited energy interval around the Fermi surface. However,
there is no recipe to determine this energy band in a unique way. It
is obvious that the density $\rho$ cannot be integrable if solutions
belonging to the continuum part of the spectrum of the Eq. (32) are
included (i.e. when the pairing field is nonvanishing for such
states). Furthermore, if one considers that $\Delta$ is acting for all
energies, then the density $\Phi$ is
\begin{eqnarray}
\Phi (x,y)& = &\sum_i \phi_i (x) \phi^*_i (y) 
\frac{\Delta}{2 \sqrt{(\varepsilon_i - \lambda)^2 + \Delta^2}} ,
\nonumber \\
&= &\frac{\Delta}{2} \sum_i \phi_i (x) \phi^*_i (y) 
\left [ 
\frac{1}{\sqrt{(\varepsilon_i - \lambda)^2 + \Delta^2} } 
- \frac{1}{\varepsilon_i-\lambda   } 
\right ]
- \frac{\Delta}{2} G (x,y,\lambda) ,\nonumber
\end{eqnarray}
where $G(x,y,\lambda)$ stands for the single particle Green function
of the Eq. (32). As is well known, the Green function diverges like
$\mid \bbox{r} - \bbox{r}^\prime \mid^{-1}$ when $\bbox{r} \rightarrow
\bbox{r}^\prime$.  Therefore, a finite density $\Phi$ cannot be defined in
this case because of this divergence. One notices that the
divergence is not logarithmic, as is usually stated in textbooks
\cite{ABM,JB}.  A local pairing field 
corresponds to a zero range two--body
interaction. Then, as one can easily show by using Eq. (31),
the density $\Phi$ will always be singular for coinciding
arguments. For the entire HFB procedure to be meaningful,
the two--body interaction, which is responsible for
pairing, must have a finite range.

\subsection{Square well single particle potential and square well pairing 
potential}

We assume that
\begin{eqnarray}
U (r)& =& U_0 \theta (R-r)  ,  \quad U_0 < 0,  \nonumber \\
\Delta (r)& = &\Delta_0 \theta (R-r), \nonumber
\end{eqnarray}
and look for solutions of Eqs. (11--12) with zero orbital momentum $l = 0$.

For $r < R$ the solution reads
\begin{eqnarray}
v_{in} (r)& =& C_+ \sin  k_+ r + C_- \sin k_- r , \nonumber \\
u_{in} (r)& =& \beta_+ C_+ \sin  k_+ r + \beta_- C_- \sin k_- r  ,
\nonumber
\end{eqnarray}
where
\begin{eqnarray}
\beta_\pm = \frac{E \mp \sqrt{E^2 - \Delta^2_0} }{\Delta_0}, \quad
\beta_+ \beta_- = 1 \nonumber
\end{eqnarray}
and $C_\pm$ stand for some constants, which have to be determined from
matching the interior with the exterior solutions and normalization.

For $r > R$ 
\begin{eqnarray}
&&v_{out} (r)= v_0 \exp (-k_0 r) ,\quad {\rm if} \quad 
\frac{\hbar^2}{2m} k^2_0 = 
- (E+ \lambda) >0  , \nonumber \\
&& u_{out} (r)= u_0 \exp (-ik_0^\prime r) \quad {\rm if} \quad 
\frac{\hbar^2}{2m} k^{\prime2}_0 = 
- (-E + \lambda) > 0 \nonumber
\end{eqnarray}
or 
\[ u_{out} (r) = u_1 \sin k^\prime_0 r + u_2 \cos k^\prime_0 r \]
if 
\[ \frac{\hbar^2}{2m} k^{\prime2}_0 = -E + \lambda > 0 \]
where $v_0 , u_0 , u_1 $ and $u_2$ have to be determined from matching
and normalization.

If $\lambda < E < -\lambda$, the spectrum is discrete and the energies
have to be determined from the matching condition
\[ 
\frac{\beta_+ (k_- \cos k_- R + k_0 \sin k_- R) (k_+ \cos k_+ R + 
  k^\prime_0 \sin k_+ R)
}{\beta_- (k_- \cos k_- R + k^\prime_0 \sin k_- R) (k_+ \cos k_+ R + 
  k_0 \sin k_+ R)}=1.
\]
If $E<\lambda$ , then one deals with a state lying in continuum. In
contrast to the case of constant pairing, one now has $ [ h,\Delta]
\neq 0$ and sufficiently deeply bound hole states acquire width.

Inside the potential well, the two components of the s.qp.w.f. form
a superposition of two s.p.w.f. with energies equal approximately
to $E + \lambda + U_0$ and $-E + \lambda + U_0$, respectively (as a
rule $\Delta_0$ is very small and can be neglected in the
determination of $k_\pm$).

In the vicinity of a hole state (the left hand side of 
Fig. 1) $C_+$ has a zero and one can show that 
\[  \frac{v_{in} (r)}{u_{in} (r)} \cong \beta_+ \cong 
    \frac{2E}{\Delta_0} 
\cong \sqrt{\frac{n_{BCS}}{1- n_{BCS}}} , \quad n_{BCS} \cong 1, \]
similar to the BCS approximation. This relation holds only inside the 
potential well and it is not valid outside it.
Far from the resonance this ratio becomes
\[ \frac{u_{in} (r)}{v_{in} (r)} \cong \beta_+ \cong \frac{2E}{\Delta_0}
\cong \sqrt{\frac{1 -n_{BCS}}{n_{BCS}}} , \quad n_{BCS} \ll 1 .\] 
Unlike the BCS approximation, the radial behaviors of the $v$-- and 
$u$--components are no longer identical.

In this case, the density $\rho$ has the correct asymptotic behavior, but 
due to the local character of the pairing potential,
the density $\Phi$ has the same divergence as in the case of BCS 
approximation. .

\subsection{Surface pairing $\Delta =\Delta _0 \delta(r-R) $ }

This is another approximation which has been used in nuclear physics. The 
solution of Eqs. (11--12) is now
\begin{eqnarray}
&&v (r) = \Delta_0 G (r,R, E+\lambda) u (R) , \nonumber \\
&& u(r) = \kappa  u_0 (r) - \Delta^2_0 G (r,R, -E+\lambda) G (R,R,E+\lambda) u
(R), \nonumber 
\end{eqnarray}
where 
\[ u (r) = \frac{\kappa  u_0 (R)}{1+\Delta^2_0 G(R,R,-E+\lambda) G(R,R, 
E+\lambda)}, \]
$\kappa$ is a normalization constant and $u_0$ is given by the
equation
\[ (h-\lambda+E) u_0 = 0 \]
($u_0$ has to be included only for $E <\lambda$)

The density $\rho$ has a correct asymptotic behavior, but the same
problems arise  with the density $\Phi$ as
above. The density of the pair condensate $\Phi$ is singular for 
$\bbox{r} =\bbox{r}^\prime$ (see Eq. (31)).

Even though the examples discussed here have little in common with the real
selfconsistent solution of Eqs. (11--12), in our opinion however, they
lead to a deeper understanding of the structure of the HFB equations in
the case of finite systems. Especially instructive in this sense is
the role played by the nonlocality of the pairing field and
consequently by the range of the two--body forces.

\section{Conclusion}

We have examined the HFB approximation in the case of finite systems, 
when the two--body interaction between particles has a finite
range. Special attention was paid to the asymptotic behavior of the
s.qp.w.fs. It was shown that the s.qp. states located in spectrum sufficiently 
far from the Fermi surface have the character of continuum
states. E.g. the deep hole states acquire a width
corresponding to the decay into a particle state and the pair
condensate. This width has to be interpreted as a contribution to the
imaginary part of the s.p. optical potential.

Even though most of the s.qp.w.fs. lie in continuum, the
matter distribution is nevertheless finite. The same 
holds true for the density of the pair condensate, which is
finite as well.

These features of the general solutions of the HFB equations have to be
included in any HFB calculations. To what extent the available HFB
results \cite{HJM,DG} correspond to the real solution, is still an
issue which needs further studies.


\begin{thebibliography}{99}

\bibitem{NNB} N.N. Bogoliubov, Usp. .Fiz. Nauk {\bf 67}, 549 (1959)
(transl. Soviet Phys. Usp. 2, 236 (1959)).

\bibitem{STB} S.T. Belyaev, Mat. Fys. Medd. Dan. Vid. Selsk. {\bf 31}, 11 (1959).

\bibitem{ABM} A.B. Migdal, {\it Theory of Finite Fermi Systems}, 
J.Wiley, N.Y. 1967.

\bibitem{JGV} J.G. Valatin, in {\it Lectures in Theoretical Physics IV},
eds.  W.E. Brittin, B.W. Downs, J. Downs, Interscience Pub., J. Wiley
and Sons, N.Y., London, 1962.

\bibitem{MB} M. Baranger, in {\it 1962 Cargese Lectures in Theoretical
Physics}, W.A. Benjamin, Inc., N.Y, Amsterdam, London, 1969.

\bibitem{ABM2} A.B. Migdal, {\it Fermions and Bosons in Strong Fields}, 
Nauka, Moscow, 1978.

\bibitem{CM} C. Mahaux, H.A. Weidenm\"uller, {\it Shell-Model Approach to
Nuclear Reactions}, North--Holland Pub. Comp., Amsterdam, London, 1969.

\bibitem{HJM} H.J. Mang, Phys. Rep. {\bf C 18}, 325 (1975).

\bibitem{DG} D. Gogny, in {\it Nuclear Self-Consistent Fields},
eds. G. Ripka, M. Porneuf, North-Holland Pub. Comp., Amsterdam, Oxford
Inc., N.Y, 1975, p. 333.

\bibitem{JB} J. Bardeen, L.N. Cooper, J.R. Schrieffer, Phys. Rev. {\bf 108},
1175 (1957).

\end{thebibliography}
\end{document}